\documentclass{amia}
\usepackage{graphicx}
\usepackage{amssymb}
\usepackage{colonequals}
\usepackage{textcomp}
\usepackage{amsmath}
\usepackage{amsthm}
\usepackage{pgf,tikz}
\usetikzlibrary{matrix, positioning, fit}
\usepackage{pgfplots}
\usetikzlibrary{shapes,arrows,snakes,automata,backgrounds,trees,calc}
\usepackage{wrapfig}
\usepackage{boxedminipage}
\usepackage{algpseudocode}
\usepackage{wrapfig}
\usepackage{subfig}
\usepackage{caption}
\usepackage{colortbl}
\usepackage{multirow}
\usepackage{url}
\usepackage{bm}

\newsavebox\curwrapfig
\makeatletter
\long\def\wrapfiguresafe#1#2#3{%
  \sbox\curwrapfig{#3}%
  \par\penalty-100%
  \begingroup 
    \dimen@\pagegoal \advance\dimen@-\pagetotal 
    \advance\dimen@-\baselineskip 
    \ifdim \ht\curwrapfig>\dimen@ 
      \break%
    \fi%
  \endgroup%
  \begin{wrapfigure}{#1}{#2}%
    \usebox\curwrapfig%
  \end{wrapfigure}%
}
\makeatother

\begin{document}

\title{TELII: Temporal Event Level Inverted Indexing for Cohort Discovery on a Large Covid-19 EHR Dataset}

\author{Yan Huang PhD}

\institutes{
    McGovern Medical School, 
    Texas Institute for Restorative Neurotechnologies\\
     The University of Texas Health Science Center at Houston, Houston, Texas 77030, USA\\
}

\maketitle

\section*{Abstract}
\textit{
  Cohort discovery is a crucial step in clinical research on Electronic Health Record (EHR) data. Temporal queries, which are common in cohort discovery, can be time-consuming and prone to errors when processed on large EHR datasets. In this work, we introduce TELII, a temporal event level inverted indexing method designed for cohort discovery on large EHR datasets. TELII is engineered to pre-compute and store the relations along with the time difference between events, thereby providing fast and accurate temporal query capabilities. We implemented TELII for the OPTUM\textsuperscript{\textregistered} de-identified COVID-19 EHR dataset, which contains data from 8.87 million patients. We demonstrate four common temporal query tasks and their implementation using TELII with a MongoDB backend. Our results show that the temporal query speed for TELII is up to 2000 times faster than that of existing non-temporal inverted indexes. TELII achieves millisecond-level response times, enabling users to quickly explore event relations and find preliminary evidence for their research questions. Not only is TELII practical and straightforward to implement, but it also offers easy adaptability to other EHR datasets. These advantages underscore TELII's potential to serve as the query engine for EHR-based applications, ensuring fast, accurate, and user-friendly query responses.
}

\section{Introduction}
\label{intro}

Secondary use of Electronic Health Record (EHR) data is a valuable resource for clinical research because its easy management of rights and consents, and its full clinical content~\cite{coorevits2013ehr}. EHR data contains a wide range of clinical information, including patient demographics, diagnoses, procedures, medications, lab tests, and vital signs. EHR data can leverage research and support critical decisions about effectiveness of drugs and therapeutic strategies.
EHR data is used to support various clinical research questions, such as disease risk factors, disease progression, treatment effectiveness and and therapeutic strategies. The foundation of these research questions is to study the relations between clinical records based on large volume of existing cases. The recent COVID-19 pandemic has underscored the urgency for extensive clinical research on EHR data, given the swift propagation and development of the epidemic, coupled with the constrained time frame research institutions have for data collection and analysis~\cite{dagliati2021covid}. 

To initialize an EHR-based study, an important step is to identify a cohort of patients who match certain criteria. The criteria can be existing conditions of the patients, such as hypertension, diabetes, cancer history, or recent vaccine records. The criteria can also be the temporal relations between events, such as find patients who had

\[
 \begin{array}{l} 
    \mbox{\em Severe symptoms of COVID-19 within 30 days after the positive PCR test};\\
    \mbox{\em Pregnancy after the first dose of COVID-19 vaccine};\\
    \mbox{\em ICU admission on the same day of the diagnosis of COVID-19};\\
    \mbox{\em COVID-19 diagnosis before readmission to hospital}.
\end{array}\]

While a variety of tools exist to aid researchers in filtering populations based on attributes or facets (for instance, i2b2~\cite{i2b2} and BTRIS~\cite{btris} in the medical field), many cohort studies necessitate their subjects to demonstrate a temporal pattern for eligibility. Regrettably, these conventional query tools often fall short in providing efficient temporal queries~\cite{cohort_selection}. The crucial yet challenging step of selecting cohorts with temporal constraints from databases is often done by writing extraction scripts, which are time-consuming, un-reusable, and error-prone. For large EHR datasets, existing algorithms and databases requires a huge amount of computational resources and time to process the temporal queries.

\begin{figure}[h]
  \centering
  \includegraphics[width=1\textwidth]{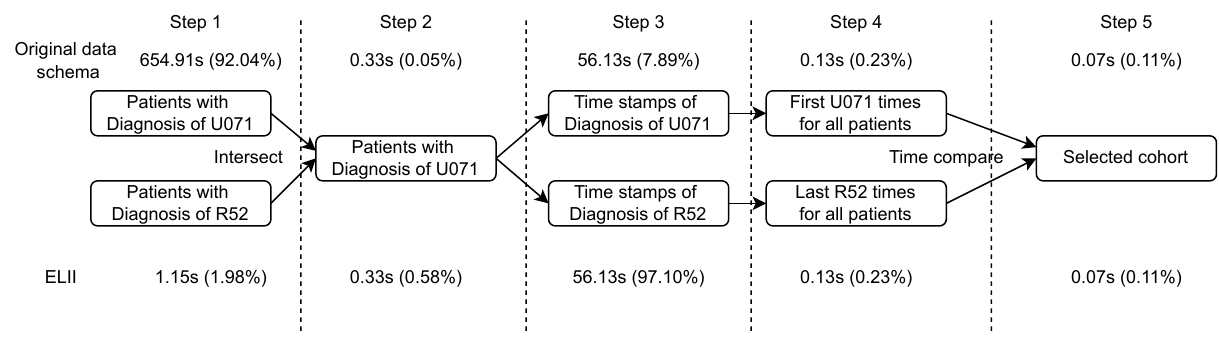}
  \caption{Computational time consumption for query ``Find patients who had Diagnosis of U071 COVID-19 before Diagnosis of R52 Pain'' by using record-based data schema or ELII. }
  \label{fig:query}
  \end{figure}

A temporal query in EHR is a data retrieval request consists of a set of clinical records and the temporal constraints between the records. The most common definition of temporal constraints is Allen's interval algebra~\cite{allen1983maintaining}. Allen's interval algebra is a set of 13 relations between two intervals, such as before, contain, overlap. Since the records in EHR are time points instead of continuous intervals, the temporal constraints between the EHR records can only be before, after, and co-occur(equal) in terms of Allen's interval algebra. To determine if a relation satisfies the query, we need to compute the time difference between the two events. 
Figure~\ref{fig:query} shows the strategy for query ``Find patients who had Diagnosis of U071 COVID-19 before Diagnosis of R52 Pain''. The query includes five steps to identify the patients with both diagnoses and satisfy the ``before'' relation between them. By using the original EHR record-based schema, the majority of the time (92.04\%) is spent on the first step when identify the patients with the diagnosis of U071 and the patients with the diagnosis of R52.

Existing works had been done to improve query performance generally. One approach is to sample the dataset to reduce the volume of the data, for example, drawing and materializing random samples from the data table rows~\cite{sample1,sample2,sample3}. The other approach is to pre-compute the expected query results or building a intermediate data structure to support the query~\cite{trie,pre-compute}. In our previous work, we proposed an a novel approach called Event Level Inverted Index (ELII) for cohort discovery on EHR data~\cite{elii}. ELII optimizes time trade-offs between one-time batch preprocessing and open-ended, user-specified temporal queries. We implemented query engine in a NoSQL (not only SQL) database using the ELII strategy to improve temporal query performance on EHR data. ELLI solves one problem of temporal query that it is inefficient to locate the patients with specific events in a large database. Figure~\ref{fig:query} shows that ELII reduces the time of the first step from 654.91 seconds to 1.15 seconds for this particular case. However, ELII does not contain the temporal info so step three is still time consuming.

In this study, we introduce TELII, a temporal event level inverted indexing for database, as an optimization for the third step in cohort discovery on large EHR datasets. TELII offers a more efficient, scalable, and flexible design compared to existing non-temporal solutions.
TELII is engineered to pre-compute and store the relations and time differences between events. It employs an inverted index for quick access to the patient list based on query inputs such as "Event A before Event B". We have built TELII for the OPTUM\textsuperscript{\textregistered} COVID-19 data and demonstrated the performance improvement from ELII to TELII. We also showcase four common temporal query tasks and their implementation using TELII with a MongoDB backend~\cite{mongodb}.
Our results reveal that TELII's temporal query speed is up to 2000 times faster than that of existing non-temporal inverted indexes. TELII achieves millisecond-level response times, enabling users to swiftly explore event relations and find preliminary evidence for their research questions. The results also highlight TELII's capability to perform preliminary statistical analysis on large EHR datasets within seconds. TELII is not only practical and straightforward to implement, but it also offers easy adaptability to other EHR datasets. These advantages underscore TELII's potential to serve as the query engine for EHR-based applications, ensuring fast, accurate, and user-friendly query responses.

\section{Methods}\label{sec:methods}
\begin{figure}[h]
\centering
\includegraphics[width=0.9\textwidth]{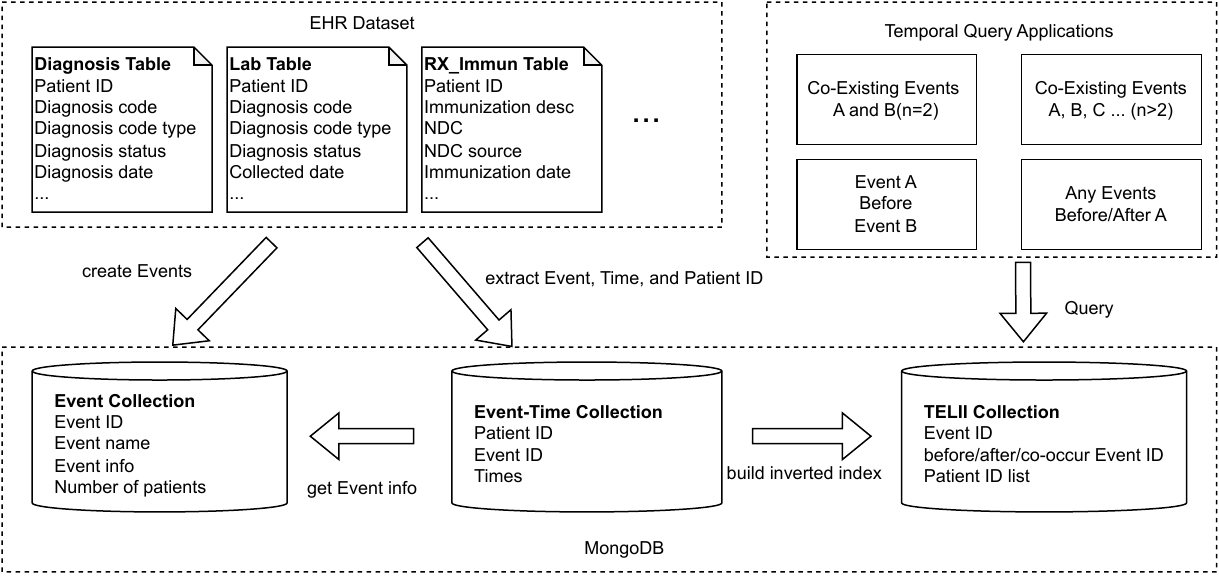}
\caption{The process of building TELII for an EHR dataset.}
\label{fig:build_telii}
\end{figure}

In this section, we describe how we build a MongoDB based TELII for an EHR dataset using the example of OPTUM\textsuperscript{\textregistered} COVID-19 data. Figure~\ref{fig:build_telii} shows the overview of TELII. We first describe the process of building three collections in MongoDB: Event, Event-Time, and the indexing collection, and then describe how to perform four different common temporal query tasks using TELII.

\subsection{Event Relation Extraction}
We used OPTUM\textsuperscript{\textregistered} COVID-19 data which is drawn from dozens of healthcare providers in the United States, including more than 700 hospitals and 7,000 clinics. In the January 2022 release, the dataset consisted of 8.87 million unique individuals who have documented clinical care with a documented diagnosis of COVID-19 or acute respiratory illness after 02/01/2020 and/or documented COVID-19 testing regardless of their results. The data incorporated a wide swath of raw clinical data, including new, unmapped COVID-specific clinical data points from both inpatient and ambulatory electronic medical records, which included patient-level information: demographics, diagnoses, procedures, lab tests, care settings, medications prescribed or administered, and mortality. These data are certified as de-identified by independent statistical experts following Health Insurance Portability and Accountability Act (HIPAA) statistical de-identification rules and are managed by the customer data use agreement. 17 individual source files were contained in the release, for different types of EHR records, such as patient demographics, diagnosis, medication, and lab. Each source file came with several types of attributes. For instance, the DIAGNOSIS source file contained attributes such as diagnosis code, diagnosis code type, and diagnosis status. Definition of these attribute types are given in the accompanying data dictionary provided by data provider.

In this work, we designed ``Event'', the query unit, to represent original EHR records by extracting most common fields for querying, such as diagnosis code, procedure code, NDC for medication, and lab test code. For example, a diagnosis event is consisted of diagnosis code, diagnosis code type (ICD-10,ICD-9, and SNOMED CT), and diagnosis status (``Diagnosis of'',``Possible diagnosis of'', and ``History of''). Such schema can be modified before building the TELII based on the specific research question. For instance, if ``primary diagnosis'' is of interest, the ``primary diagnosis'' attribute can be added to the diagnosis event. We also pre-defined some ``Events'' that are not directly available in the dataset but useful to the research topics, such as ``COVID-19 PCR test with positive result''. To create this event, we used all the combination of lab test codes and lab test names related to COVID-19 PCR test and the result texts that indicate positive result. When user want to query for ``patients with COVID-19 confirmed by PCR test'', TELII will use this pre-processed event.

During the process of building events, we also counted the number of patients for each event. The number of patients for each event tells how common the event is in the dataset. This information is especially useful for TELII to determine the anchor event for a temporal query which is the key optimization for querying multiple events with large numbers of cases.

With all the events created and stored in Event collection, we can transform the original EHR records into combinations of Patient ID, Event ID, and Event Time. Here, the Event ID is the unique integer for each event created by the order of its number of patient.  The larger the number of patients for an event, the smaller the Event ID. We group the time of events by patient and event, and sort the times in an array. In the MongoDB Event-Time collection, one event with one or more time points of a patient is stored as one document. The document is in the format of \{``PatientID'':``PTxxxxxxx'', ``EventID'':``xxx'', ``Times'': [``date 1'', ``date 2'', ... ,  ``date n'']\}.

After the Event-Time collection is built, we can extract event relations from the collection. For an anchor event A, we define three types of relations (before, after, and co-occur) for another Event B that exists in the timeline of same patient. In MongoDB, \{ ``EventID'': ``A ID'', ``before'': ``B ID''\} means Event B happened before Event A. \{ ``EventID'': ``A ID'', ``after'': ``B ID''\} means Event B happens after Event A. \{ ``EventID'': ``A ID'', ``co-occur'': ``B ID''\} means event A and event B happen in the same date. Since co-occur relation is a special case of before and after relation when time difference equals to zero, in this work, the inverted index for before and after relation also includes the co-occur relation.
The computation of event-relation extraction is patient independent. We use parallel computing to calculate the event relations for each patient. The computation is done in the following steps: 1) for each patient, group the events by time to get co-occurrence relations  2) get and sort the first and the last time points for all events, scan the sorted time points to calculate the before relations; 3) store the temporary results in the MongoDB as the input of inverted indexing computing.

\subsection{Inverted Index using Event Relations}
The inverted index is designed to store all the existing event pairs in the dataset with a list of patients who have the event pairs.
We use one MongoDB collection to store the inverted index and each document in the collection stores an anchor event , a related event, the relation to the anchor event, and a patient ID list. For example, the inverted index for Event B before Event A is in the format of \{``EventID'':``A ID'', ``before'':``B ID'', ``PatientIDList'': [``PTxxxxxxx'', ``PTxxxxxxx'', ... ,  ``PTxxxxxxx'']\}. For each relation, the anchor event is determined by their appearance frequency in the dataset. The less common event is the anchor event of an event pair.
 Use the anchor event in the inverted index has the following advantages: 
1) all queries with same meaning will use the same inverted index. For instance, assume A is less common than B, then ``A before B'' and ``B after A'' will be translated to query \{``EventID'': ``A ID'', ``after'': ``B ID''\}, and ``A co-occur B'' and ``B co-occur A'' will be translated to query \{``EventID'': ``A ID'', ``co-occur'': ``B ID''\}.
2) instead of using one large index (Event A, Event B, relation), the collection is split into three small indices (Event ID, before), (Event ID, after), and (Event ID, co-occur). This design is more efficient for query processing.
3) the design reduces the indexing size for common events.

To support temporal queries with time difference, such as ``Event A before Event B with time difference less than 30 days'', we design a more precise inverted index. For anchor Event A and its related Event B, the format of the inverted index is \{``EventID'':``A ID'', ``RelatedEventID'':``B ID'', ``TimeDifference'': 30, ``PatientIDList'': [``PTxxxxxxx'', ``PTxxxxxxx'', ... ,  ``PTxxxxxxx'']\}. Such design allows TELII to directly return the patient list that satisfies the time difference condition without further computation. However, the calculation of time difference is more time consuming than the regular inverted index, and it needs much more storage space. 

\subsection{TELII Temporal Query with MongoDB}
We list four common temporal query tasks and their implementation using TELII with MongoDB backend. 

{\bf Co-existing of two events}. The first query task is to find patients who have two specific events in their EHR records.
For example, we want to find patients with both COVID-19 PCR test with positive result(EventID=423) and diagnosis of I10 essential (primary) hypertension(EventID=108). To use TELII to perform the query, we need to search both before and after relations for the two events. Because I10 essential (primary) hypertension diagnosis is more common than COVID-19 PCR test with positive result, we use I10 as the anchor event. The query is translated to MongoDB find statement \{ ``or'': [\{``EventID'': 108, ``before'': 423\}, \{``EventID'': 108, ``after'': 423\}]\}. 

{\bf Co-existing of a event group}. The second query task is to find patients who have a group of specific events (n$>$2) in their EHR records. For example, we want to find patients with both COVID-19 PCR test with positive result(EventID=423), diagnosis of I10 essential (primary) hypertension(EventID=108), and diagnosis of R52 Pain(EventID=659). Without TELII, we need to retrieve three large separate patient list and perform intersection. With TELII, we can use the anchor Event, R52 in this case, to get the smaller patient lists for the co-existing (task 1) of 1) R52 and I10, and 2)R52 and PCR positive. Then we perform intersection of the two patient lists to get the final result.

{\bf Before}. The third query task is to find patients who have an event A before an event B. This is the most simple and efficient query task in TELII. The task only runs one MongoDB find query \{``EventID'': ``A ID'', ``after'': ``B ID''\} or \{``EventID'': ``B ID'', ``before'': ``A ID''\}. And the result is one MongoDB document with the patient list.

{\bf Event relation exploring}. The fourth query task is to extract all Events before/after/co-occur with the input Event. For example, we want to know the most common diagnosis after COVID-19 PCR test with positive result. The task runs the MongoDB query \{ ``or'': [\{``EventID'': 423, ``after'': \{ ``\$exists'': True\}\}, \{``before'': 423\}]\} and then group the result by the related Event ID and count the number of patients.

\section{Experimental Results}
In this section, we show the performance of TELII for COVID-19 data released in January 2022. The database is hosted on a IBM server with ppc64le architecture, 128-core POWER8 CPU and 1TB memory. Operating system is Red Hat Enterprise Linux Server release 7.8, and the MongoDB version is 5.0.2. The client machine is a 2019 Mac Pro with a 2.5 GHz 28-Core Intel Xeon W processor and 1.5TB 2933 MHz DDR4 memory. The operating system is macOS Ventura version 13.6.3. We used Python3.8 to implement the TELII query tasks. We use parallel computing only for creating TELII, the query tasks were run on a single thread and the MongoDB is not distributed for the testing.

The dataset contains 8.87 million subjects and we extract 1,197,051 unique events from Diagnosis, Lab Test, Observation, Procedure, Administered Medication, Immunization, Patient Reported Medication, and Prescribed Medication. The average number of patients per event is 2,621. The most common event is the observation of weight with 7,085,345 patients, and the most common diagnosis is Z23 Encounter for immunization with 3,362,322 patients.
We created a pre-defined event ``COVID-19 PCR test with positive result'' with 996,645 patients. In this experiment, we focused on the diagnosis events related to ``COVID-19 PCR test with positive result''.
Based on the four temporal query tasks, we selected six testing events: Diagnosis of a) I10 essential (primary) hypertension (number of patients=2,569,555), b) R05 Cough (number of patients=1,991,707), c) J02.9 Acute pharyngitis, unspecified (number of patients=1,486,795), d) R53.83 Other fatigue (number of patients=1,262,188), e) R52 Pain (number of patients=669,324), f) R05.2 Subacute cough (number of patients=559). Event a-e are common diagnosis and their related queries are expected to have large number of patients and time consuming. Event f is a less common diagnosis and its related queries are expected to have small number of patients and fast response time.

We build four query methods for the query performance testing. The first method is using the original source data in SQL table format. During the experiment, we found that it is inefficient to perform testing queries without any optimization, so we did not show its results. The second method (ELII-DIAG) is using ELII with only diagnosis event imported. The third method (TELII-DIAG) is using TELII with only diagnosis event imported. The fourth method (TELII) is using TELII with all source data imported. The TELII volume of the documents in method four is about 17 times larger than the TELII in method three. From the results, we can see the performance of method three and four is similar, which shows the scalability of TELII on the large dataset. 
The query performance is measured by the response time of the query tasks. We define method A is X times faster than method B where X = (response time of method B - response time of method A)/(response time of method A). Here we list the query performance comparison for each query task.

\begin{figure}[h]
  \centering
  \includegraphics[width=0.9\textwidth]{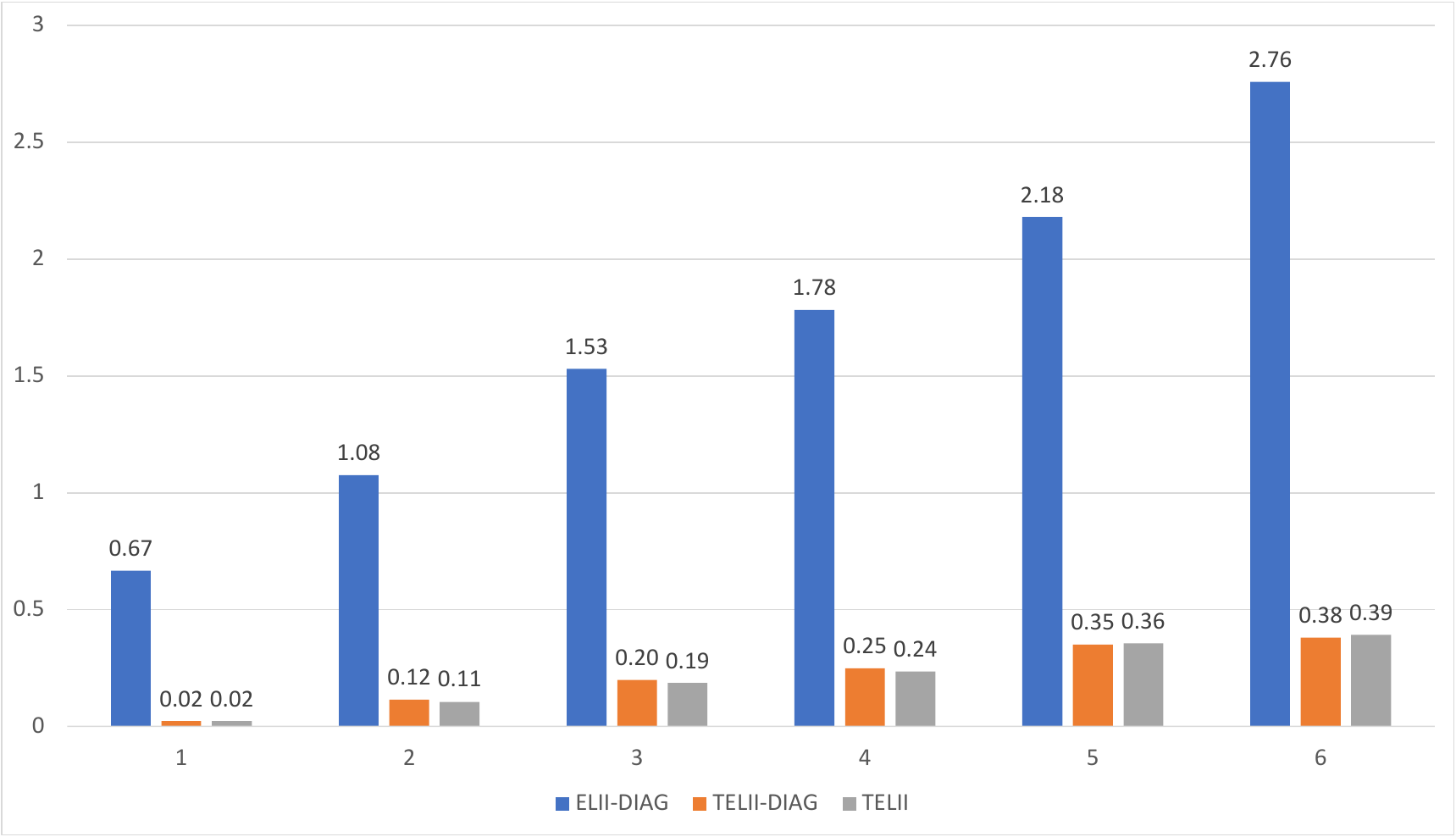}
  \caption{Result1. The query performance of ELII-DIAG, TELII-DIAG, and TELII on testing task 1 query 1-6. The response time is in seconds.}
  \label{fig:result1}
\end{figure}

{\bf Result1: Co-existing of two events}. We performed six testing queries: Find patients with both 1) ``COVID-19 PCR positive'' and R05.2 Subacute cough, 2) ``COVID-19 PCR positive'' and R52 Pain, 3) ``COVID-19 PCR positive'' and R53.83 Other fatigue, 4) ``COVID-19 PCR positive'' and J02.9 Acute pharyngitis, 5) ``COVID-19 PCR positive'' and R05 Cough, 6) ``COVID-19 PCR positive'' and I10 essential (primary) hypertension.
The query performance is shown in Figure~\ref{fig:result1}. The query speed for TELII-DIAG is 10.25 times faster than ELII-DIAG in average. The queries were sorted by the number of patients for the related event from small to large. With the increase of the number of patients, the response times of both ELII-DIAG and TELII-DIAG increase linearly, but the increase slop of TELII-DIAG is much smaller than the slop of ELII-DIAG.

\begin{figure}[h]
  \centering
  \includegraphics[width=0.9\textwidth]{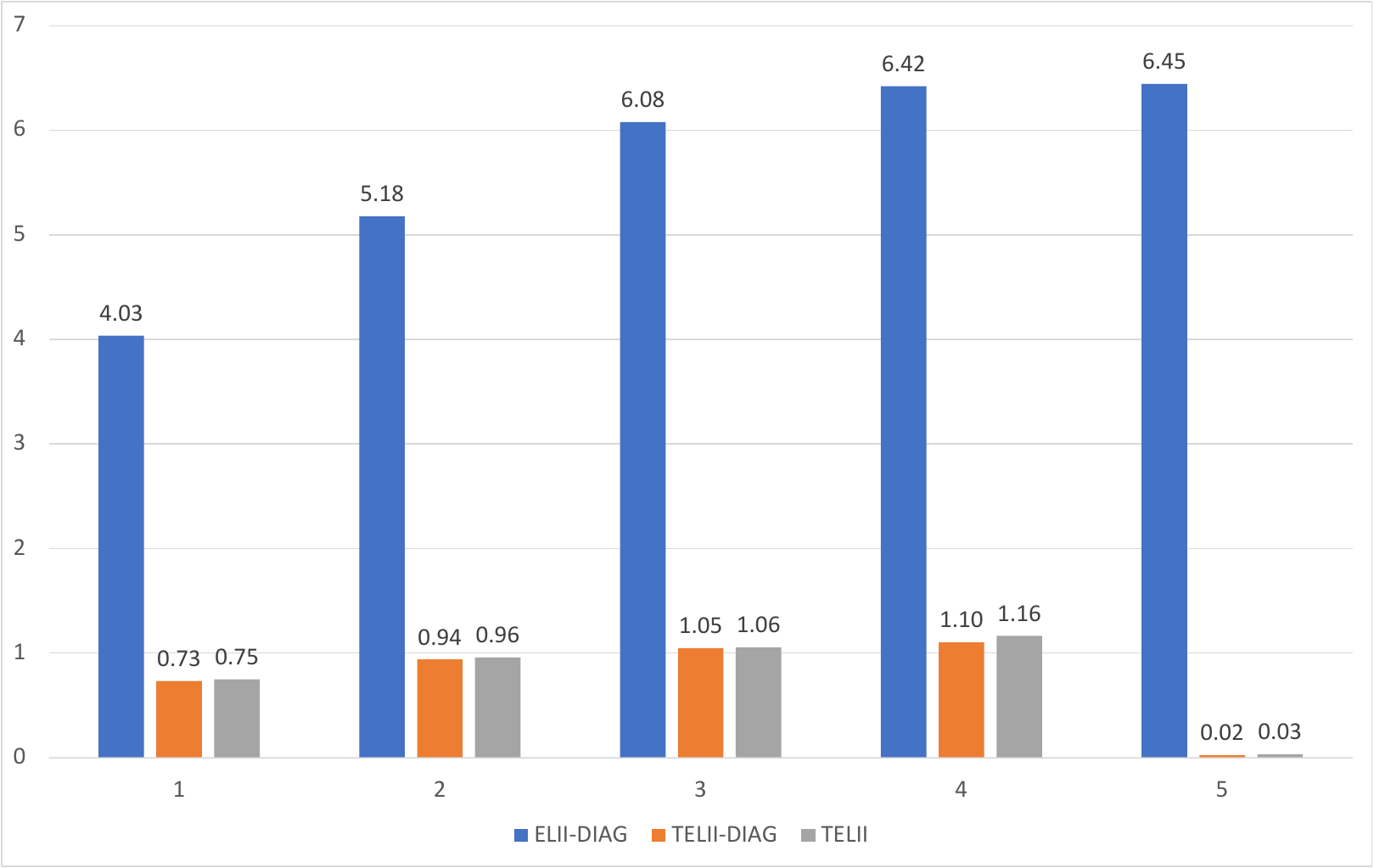}
  \caption{Result2. The query performance of ELII-DIAG, TELII-DIAG, and TELII on testing task 2 query 1-5. The response time is in seconds.}
  \label{fig:result2}
\end{figure}

{\bf Result2: Co-existing of a event group}. We performed five testing queries: Find patients with 1) three events (``COVID-19 PCR positive'', I10 essential (primary) hypertension, and R05 Cough), 2) four events (``COVID-19 PCR positive'', I10 essential (primary) hypertension, R05 Cough, and J02.9 Acute pharyngitis), 3) five events (``COVID-19 PCR positive'', I10 essential (primary) hypertension, R05 Cough, J02.9 Acute pharyngitis, and R53.83 Other fatigue), 4) six events (``COVID-19 PCR positive'', I10 essential (primary) hypertension, R05 Cough, J02.9 Acute pharyngitis, R53.83 Other fatigue, and R52 Pain), 5) seven events (``COVID-19 PCR positive'', I10 essential (primary) hypertension, R05 Cough, J02.9 Acute pharyngitis, R53.83 Other fatigue, R52 Pain, and R05.2 Subacute cough). 
The query performance is shown in Figure~\ref{fig:result2}. The query speed for TELII-DIAG is 4.66 times faster than ELII-DIAG in average for common events group query (1-4). For query 5 which includes a less common event R05.2, the query speed for TELII-DIAG much faster (275 times) than ELII-DIAG. It shows the advantage of pre-processing on event relations by filtering out unnecessary patients for the less common event.

\begin{figure}[h]
  \centering
  \includegraphics[width=0.9\textwidth]{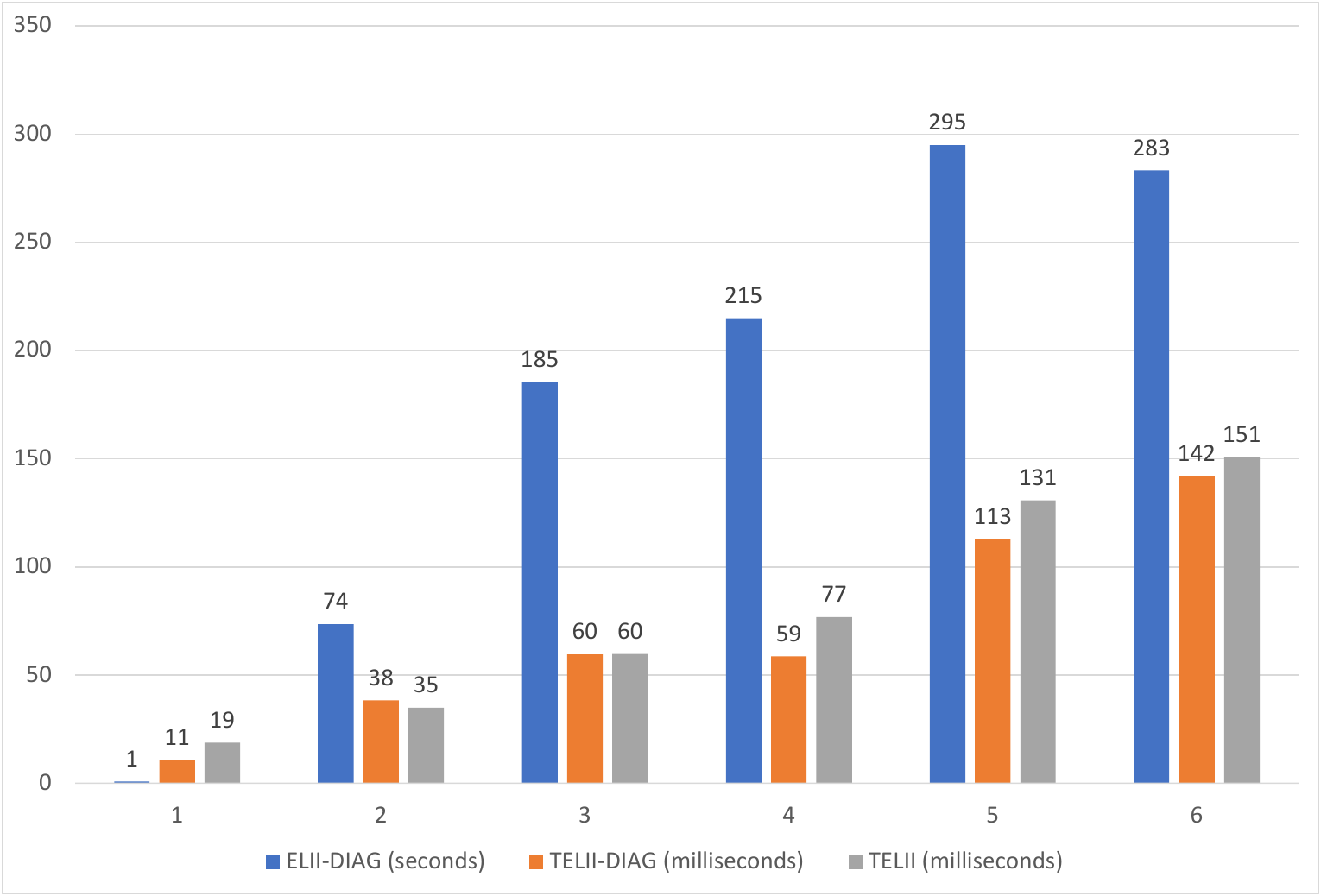}
  \caption{Result3. The query performance of ELII-DIAG, TELII-DIAG, and TELII on testing task 3 query 1-6. The response time is in seconds for ELII-DIAG, and in milliseconds for TELII-DIAG and TELII.}
  \label{fig:result3}
\end{figure}

{\bf Result3: Before}. We performed six testing queries: Find patients who have 1) ``COVID-19 PCR positive'' before R05.2 Subacute cough, 2) ``COVID-19 PCR positive'' before R52 Pain, 3) ``COVID-19 PCR positive'' before R53.83 Other fatigue, 4) ``COVID-19 PCR positive'' before J02.9 Acute pharyngitis, 5) ``COVID-19 PCR positive'' before R05 Cough, 6) ``COVID-19 PCR positive'' before I10 essential (primary) hypertension.
The query performance is shown in Figure~\ref{fig:result3}. Note that the response time for TELII-DIAG and TELII is in milliseconds. The query speed for TELII-DIAG is about 2,000 times faster than ELII-DIAG in average. The result shows that the computation of time difference on the fly is much more time consuming than using pre-computed inverted index.

{\bf Result4: Event relation exploring}. We performed four testing queries: Get all the diagnosis events that occurred after 1) ``COVID-19 PCR positive'' within 0-30 days, 2) ``COVID-19 PCR positive'' within 31-60 days, 3) J101 Influenza within 0-30 days, 4) J101 Influenza within 31-60 days. The query template were used in our previous COVID-19 vaccine study~\cite{vaccine}.
Since ELII cannot perform such queries efficiently, we only show the query results of TELII in Table~\ref{tab:result4}. The table includes the top 15 diagnosis events and their number of patients (with percentage among all patients with the given event) that the diagnoses were recorded after the given event within 0-30 days and 31-60 days. The tasks can finish in 3.43 - 19.84 seconds, which is reasonable for the large dataset. Such tasks is useful for researchers to explore the event relations and find preliminary evidence for the research questions.

\begin{table}[h]
\centering
\caption{Result 4. The query results of event relation exploring.}
\label{tab:result4}

\begin{tabular}{|c|c|c|c|c|c|c|c|c|}
\hline
\multirow{3}{*}{Query} & \multicolumn{4}{c|}{COVID-19 PCR Confirmed} & \multicolumn{4}{c|}{J10.1 Influenza} \\ \cline{2-9} 
                      & \multicolumn{2}{c|}{After within 0-30 days} & \multicolumn{2}{c|}{After within 31-60 days} & \multicolumn{2}{c|}{After within 0-30 days} & \multicolumn{2}{c|}{After within 31-60 days} \\ \cline{2-9} 
                      & \multicolumn{2}{c|}{Response time: 19.84s } & \multicolumn{2}{c|}{Response time: 9.66s } & \multicolumn{2}{c|}{Response time: 6.67s } & \multicolumn{2}{c|}{Response time: 3.43s} \\ \hline
Rank& ICD10 & {\#} Patients & ICD10 & {\#} Patients & ICD10 & {\#} Patients & ICD10 &  {\#} Patients \\ \hline
1 & J96.01 & 69,668 (6.99\%) & E11.9  & 19,227 (1.93\%)  & R50.9  & 78,748 (34.78\%)  & I10  & 12,447 (5.5\%) \\ \hline 
2 & J12.89 & 56,179 (5.64\%) & Z86.16  & 16,967 (1.7\%)  & R05  & 56,663 (25.02\%)  & Z79.899  & 5954 (2.63\%) \\ \hline 
3 & E11.9 & 537,09 (5.39\%) & E03.9  & 113,26 (1.14\%)  & J02.9  & 28,999 (12.81\%)  & R05  & 5,885 (2.6\%) \\ \hline
4 & J12.82 & 50,834 (5.1\%) & I25.10  & 10,966 (1.1\%)  & I10  & 27,334 (12.07\%)  & Z23  & 5,825 (2.57\%) \\ \hline
5 & R51.9 & 45,448 (4.56\%) & Z86.19  & 10,659 (1.07\%)  & J11.1  & 25,334 (11.19\%)  & E78.5  & 5,438 (2.4\%) \\ \hline
6 & J18.9 & 42,719 (4.29\%) & E66.01  & 10,146 (1.02\%)  & R68.89  & 21,364 (9.43\%)  & K21.9  & 5,243 (2.32\%) \\ \hline
7 & R09.81 & 42,303 (4.24\%) & D64.9  & 9,425 (0.95\%)  & Z79.899  & 18,542 (8.19\%)  & E11.9  & 4,904 (2.17\%) \\ \hline
8 & Z11.52 & 42,175 (4.23\%) & G47.33  & 9,374 (0.94\%)  & J06.9  & 16,949 (7.48\%)  & J06.9  & 4,883 (2.16\%) \\ \hline
9 & N17.9 & 39,476 (3.96\%) & Z79.4  & 9,184 (0.92\%)  & E78.5  & 14,273 (6.3\%)  & Z00.129  & 4,662 (2.06\%) \\ \hline
10 & R09.02 & 39,337 (3.95\%) & E78.2  & 8,862 (0.89\%)  & Z87.891  & 13,355 (5.9\%)  & R50.9  & 4,442 (1.96\%) \\ \hline
11 & B34.9 & 36,129 (3.63\%) & Z79.01  & 8,502 (0.85\%)  & R06.02  & 12,415 (5.48\%)  & F41.9  & 4,288 (1.89\%) \\ \hline
12 & R52 & 33,986 (3.41\%) & E11.65  & 8,398 (0.84\%)  & K21.9  & 12,355 (5.46\%)  & J02.9  & 4,104 (1.81\%) \\ \hline
13 & R05.9 & 33,053 (3.32\%) & J96.01  & 8,386 (0.84\%)  & J45.909  & 11,691 (5.16\%)  & Z87.891  & 3,515 (1.55\%) \\ \hline
14 & R94.31 & 31,875 (3.2\%) & Z79.82  & 6,805 (0.68\%)  & E11.9  & 11,180 (4.94\%)  & J44.9  & 3,395 (1.5\%) \\ \hline
15 & I25.10 & 31,458 (3.16\%) & J12.82  & 6,586 (0.66\%)  & R52  & 10,886 (4.81\%)  & J45.909  & 3,353 (1.48\%) \\ \hline

\end{tabular}
\end{table}

\section{Discussion}
The significant performance improvement of TELII is not achieved without cost. It necessitates more time for pre-processing and a substantial amount of storage space. In our experiment, which was based on the COVID-19 data, the Diagnoses-only TELII contained 1 billion documents and utilized 539GB of storage, while the full TELII contained 21 billion documents and required 21TB of storage. In comparison, the non-temporal indexed ELII, in its full form, contained 1 million documents and used 34GB of storage. On average, TELII uses more than 600 times the storage and achieves up to 2000 times the speed improvement. The balance between space and speed is a trade-off, and it's crucial to consider the specific research question and the dataset size before constructing TELII. We discovered that as the dataset size increases, including the number of patients and events, the performance of TELII remains stable. However, an increase in the number of patients results in linear growth of storage, while an increase in the number of events leads to squared growth of storage. We also found that if the number of patients for an event is relatively small (e.g., less than 10,000), the performance is acceptable (see Result3 task 1). If storage space is a concern, we recommend a hybrid solution, such as using TELII for the most common events and ELII for the less common ones. This approach will meet most requirements, such as those of a web application.

TELII can be easily adapted to other EHR datasets and can be utilized to support a variety of clinical research questions. The indexing schema of TELII is simple and reusable for any type of data with timestamps, such as EEG signal data with annotations. TELII supports ``or'' and ``negation'' logic, although these are not discussed in this paper. Furthermore, TELII has the potential to serve as a central query engine for multiple data sources by integrating ontology into the event-building process. The MongoDB query for TELII is straightforward to implement and can be modified to support tasks beyond the four temporal queries demonstrated in this paper. Currently, TELII only supports events with a single time point. In the future, we plan to extend TELII to support events with multiple time points or time intervals, enabling it to query relations in Allen's interval algebra.

TELII does have its limitations. It is not designed to handle complex temporal queries involving multiple relations, such as "find patients with event A before event B before event C". This query differs logically from "find patients with event A before event B and event B before event C", which TELII optimizes. The template "event A before event B before event C" necessitates checking if event B is the same event that occurs after event A and before event C. In TELII, this requires returning all time points for event B that satisfy both conditions and then performing an intersection. While this can be implemented in TELII, the performance may not be as efficient as it is for simpler queries. Another limitation is that TELII is not designed for real-time data insertion. When new data arrives, all existing inverted indices related to the new events need to be updated. This process is inefficient because updating is much slower than inserting in MongoDB for large collections. In our experiment, we found that the updating process takes more time than rebuilding the entire TELII. Therefore, TELII is more suitable for stable datasets or those where the update period is long enough to allow for a complete rebuild of TELII.

\section{Conclusion}
In this work, we present TELII, a temporal event level inverted indexing for cohort discovery on a large EHR dataset. TELII is designed to support millisecond level temporal queries with large numbers of events and patients. We demonstrate the performance of TELII on a large COVID-19 EHR dataset and show that TELII can achieve up to 2000 times the speed improvement over non-temporal indexed solution. We also show that TELII is scalable and can support a variety of clinical research questions. TELII is a powerful database tool for clinical researchers to explore the temporal relations of events in EHR data and to discover cohorts for clinical research.


{\small

}
\end{document}